\documentclass{article}
\usepackage{spconf,amsmath,graphicx,amsfonts}
\allowdisplaybreaks
\usepackage{regensky}

\usepackage[firstpage=true]{background}

\SetBgContents{\parbox{\textwidth}{\footnotesize © 2023 IEEE. Personal use of this material is permitted. Permission from IEEE must be obtained for all other uses, in any current or future media, including reprinting/republishing this material for advertising or promotional purposes, creating new collective works, for resale or redistribution to servers or lists, or reuse of any copyrighted component of this work in other works.}}
\SetBgScale{1}
\SetBgAngle{0}
\SetBgPosition{current page.north}
\SetBgVshift{-1.6cm}
\SetBgColor{black}
\SetBgOpacity{1}

\title{Geometry-Corrected Geodesic Motion Modeling with\\ Per-Frame Camera Motion for 360-Degree Video Compression}

\name{Andy Regensky and André Kaup\thanks{The authors gratefully acknowledge that this work has been supported by the Deutsche Forschungsgemeinschaft (DFG, German Research Foundation) under project number 418866191.}}
\address{Friedrich-Alexander-Universität Erlangen-Nürnberg\\Multimedia Communications and Signal Processing\\Cauerstr. 7, 91058 Erlangen, Germany}

\begin{document}
\maketitle

\begin{abstract}
  The large amounts of data associated with 360-degree video require highly effective compression techniques for efficient storage and distribution.
  The development of improved motion models for 360-degree motion compensation has shown significant improvements in compression efficiency.
  A geodesic motion model representing translational camera motion proved to be one of the most effective models.
  In this paper, we propose an improved geometry-corrected geodesic motion model that outperforms the state of the art at reduced complexity.
  We additionally propose the transmission of per-frame camera motion information, where prior work assumed the same camera motion for all frames of a sequence.
  Our approach yields average Bjøntegaard Delta rate savings of 2.27\% over H.266/VVC, outperforming the original geodesic motion model by 0.32 percentage points at reduced computational complexity.
\end{abstract}


\section{Introduction}\label{sec:introduction}

Virtual and augmented reality applications for entertaining, educating and connecting people impose high requirements on associated 360-degree video data.
To ensure immersive experiences and to avoid motion sickness, high resolution, framerate and visual quality are of major importance.
These requirements lead to immense amounts of data, which call for highly effective video compression to reduce costs associated with the storage and distribution of 360-degree video.

\begin{figure}
  \centering
  \subfloat[Spherical domain\label{fig:360image:spherical}]{%
  \includegraphics[width=0.32\linewidth]{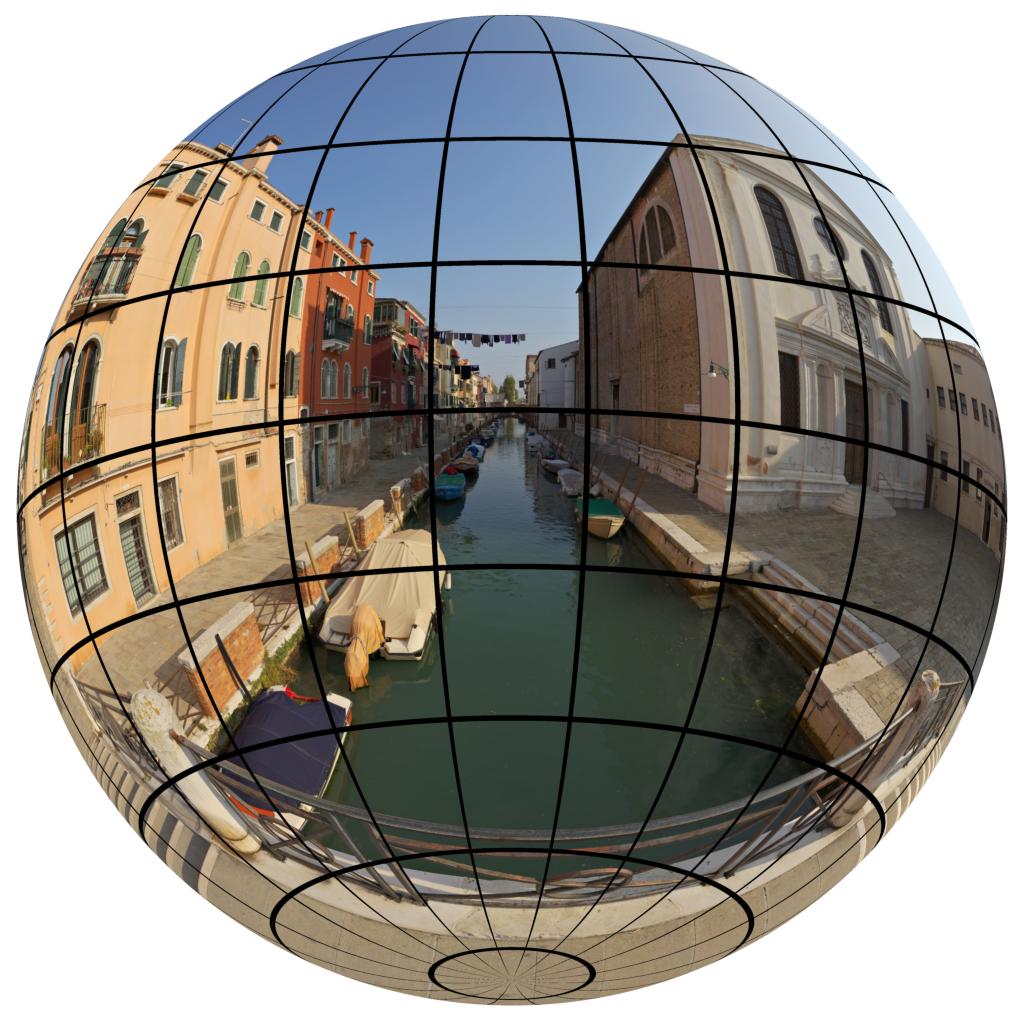}}
  \hfill%
  \subfloat[Equirectangular projection\label{fig:360image:erp}]{%
  \includegraphics[width=0.64\linewidth]{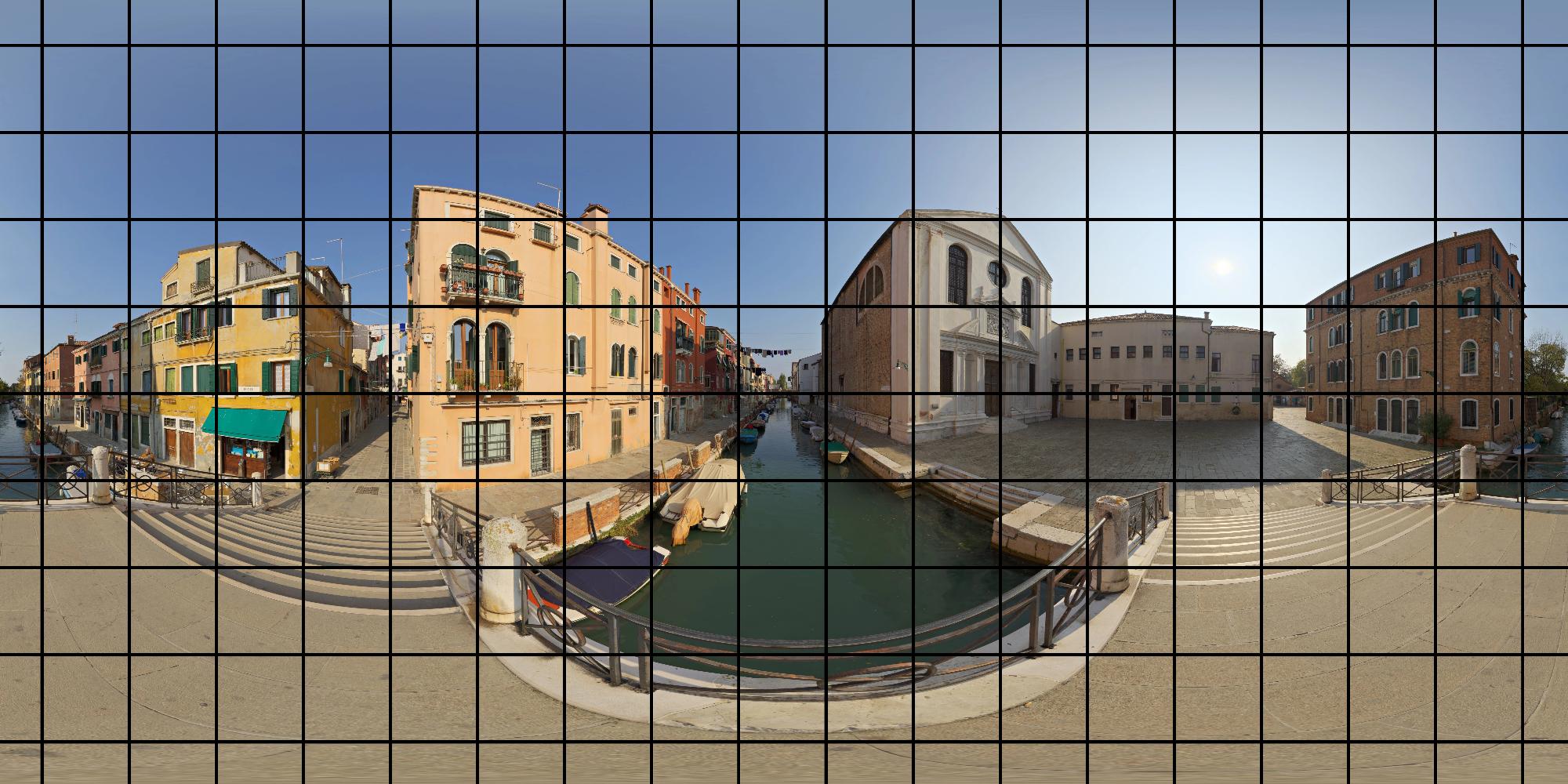}}
  \caption{Spherical video frame (a) and its representation on the 2D image plane using the equirectangular projection (b).}
  \label{fig:360image}
\end{figure}

In general, a 360-degree video provides video data for an all-around field of view and is naturally represented to lie on the surface of a sphere in 3D space as shown in Fig.~\ref{fig:360image:spherical}.
In order to allow compression using existing techniques such as the H.266/VVC~\cite{Bross2021a} video coding standard, the 360-degree video data needs to be mapped to a 2D image plane representation.
One of the most common projection formats to achieve this is the equirectangular projection shown in Fig.~\ref{fig:360image:erp}.
It maps the longitudes and latitudes of the video in 3D space to the horizontal and vertical axes of the 2D image plane, respectively.

However, existing video coding standards suffer from the inevitable distortions~\cite{Pearson1990} that occur in 360-degree video~\cite{Bross2021a, Wien2017}.
To improve upon this, many components of modern video codecs have been investigated and adapted, aiming to enhance compression efficiency for 360-degree video.
As accurate motion compensation plays a key role in overall video compression performance, many studies addressed the development and design of improved motion models for 360-degree motion compensation~\cite{Li2019, DeSimone2017, Vishwanath2017, Vishwanath2022}.
Among these, the geodesic motion model by \textit{Vishwanath et al.}~\cite{Vishwanath2022} proved to be one of the most effective.
It models translational camera motion with known motion direction through non-linear geodesic displacements in the spherical domain.

In this paper, we propose an improved geometry-corrected geodesic motion model that outperforms the current state of the art at a reduced computational complexity.
While prior works assumed the same camera motion direction for all frames of a sequence, we propose to extend this model by allowing per-frame camera motion information to be transmitted in the compressed bitstream.
We estimate per-frame camera motion information at the encoder and perform a detailed investigation of the effects that varying camera motion estimation accuracies have on the resulting compression efficiency in the H.266/VVC video coding standard.

\begin{figure*}%
  \centering%
  \hfill%
  \subfloat[Epipole-oriented coordinate system\label{fig:geometry:sphere}]{\input{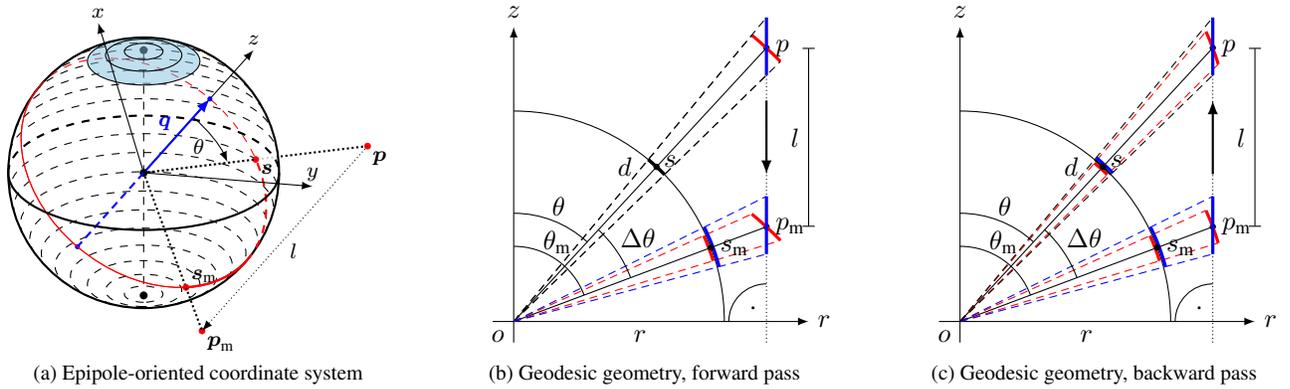}}%
  \hfill%
  \subfloat[Geodesic geometry, forward pass\label{fig:geometry:fwd}]{
\newcommand\setAngleTheta[2]{
\pgfmathsetmacro\angleTheta{atan(#2/#1)}
}

\newcommand\drawBlockRadial[3]{
\pgfmathsetmacro\blockstart{#2 - #3/2}
\pgfmathsetmacro\blockend{#2 + #3/2}
\pgfmathsetmacro\depth{#1 / cos(#2)}
\draw[very thick, black] (\blockstart:1) arc (\blockstart:\blockend:1);
\draw[very thick, red] (\blockstart:\depth) arc (\blockstart:\blockend:\depth);
\draw[densely dashed, black] (0,0) -- (\blockstart:\depth);
\draw[densely dashed, black] (0,0) -- (\blockend:\depth);
}

\newcommand\drawBlockCylinder[3]{
\pgfmathsetmacro\blockstart{#2 - #3/2}
\pgfmathsetmacro\blockend{#2 + #3/2}
\pgfmathsetmacro\blockstartY{#1 * tan(\blockstart)}
\pgfmathsetmacro\blockendY{#1 * tan(\blockend)}
\draw[very thick, black] (\blockstart:1) arc (\blockstart:\blockend:1);
\draw[very thick, blue] (#1, \blockstartY) -- (#1, \blockendY);
\draw[densely dashed, black] (0,0) -- (#1, \blockstartY);
\draw[densely dashed, black] (0,0) -- (#1, \blockendY);
}

\newcommand\drawBlockRadialMoved[4]{
\pgfmathsetmacro\blockstart{#2 - #3/2}
\pgfmathsetmacro\blockend{#2 + #3/2}
\pgfmathsetmacro\depth{#1 / cos(#2)}
\pgfmathsetmacro\yshift{-#4}
\pgfmathsetmacro\blockstartX{\depth * cos(\blockstart)}
\pgfmathsetmacro\blockstartY{\depth * sin(\blockstart) - #4}
\pgfmathsetmacro\blockendX{\depth * cos(\blockend)}
\pgfmathsetmacro\blockendY{\depth * sin(\blockend) - #4}
\pgfmathsetmacro\blockstartTheta{atan(\blockstartY/\blockstartX)}
\pgfmathsetmacro\blockendTheta{atan(\blockendY/\blockendX)}
\draw[very thick, red, shift={(0, \yshift)}] (\blockstart:\depth) arc (\blockstart:\blockend:\depth);
\draw[densely dashed, red] (0,0) -- (\blockstartX, \blockstartY);
\draw[densely dashed, red] (0,0) -- (\blockendX, \blockendY);
\draw[very thick, red] (\blockstartTheta:0.99) arc (\blockstartTheta:\blockendTheta:0.99);
}

\newcommand\drawBlockCylinderMoved[4]{
\pgfmathsetmacro\blockstart{#2 - #3/2}
\pgfmathsetmacro\blockend{#2 + #3/2}
\pgfmathsetmacro\blockstartY{#1 * tan(\blockstart) - #4}
\pgfmathsetmacro\blockendY{#1 * tan(\blockend) - #4}
\pgfmathsetmacro\blockstartTheta{atan(\blockstartY/#1)}
\pgfmathsetmacro\blockendTheta{atan(\blockendY/#1)}
\draw[very thick, blue] (#1, \blockstartY) -- (#1, \blockendY);
\draw[densely dashed, blue] (0,0) -- (#1, \blockstartY);
\draw[densely dashed, blue] (0,0) -- (#1, \blockendY);
\draw[very thick, blue] (\blockstartTheta:1.01) arc (\blockstartTheta:\blockendTheta:1.01);
}

\newcommand\drawBlockRadialMovedBack[4]{
\pgfmathsetmacro\blockstart{#2 - #3/2}
\pgfmathsetmacro\blockend{#2 + #3/2}
\pgfmathsetmacro\blockstartY{#1 * tan(\blockstart) - #4}
\pgfmathsetmacro\blockendY{#1 * tan(\blockend) - #4}
\pgfmathsetmacro\blockstartTheta{atan(\blockstartY/#1)}
\pgfmathsetmacro\blockendTheta{atan(\blockendY/#1)}

\pgfmathsetmacro\blockcenterY{#1 * tan(#2) - #4}
\pgfmathsetmacro\blockmovedTheta{atan(\blockcenterY/#1)}

\pgfmathsetmacro\depth{#1 / cos(\blockmovedTheta)}
\pgfmathsetmacro\blockstartX{\depth * cos(\blockstartTheta)}
\pgfmathsetmacro\blockstartY{\depth * sin(\blockstartTheta) + #4}
\pgfmathsetmacro\blockendX{\depth * cos(\blockendTheta)}
\pgfmathsetmacro\blockendY{\depth * sin(\blockendTheta) + #4}
\draw[thick, blue] (\blockstartTheta:\depth) arc (\blockstartTheta:\blockendTheta:\depth);
\draw[densely dotted, red] (0,0) -- (\blockstartTheta:\depth);
\draw[densely dotted, red] (0,0) -- (\blockendTheta:\depth);
\draw[thick, blue, shift={(0, #4)}] (\blockstartTheta:\depth) arc (\blockstartTheta:\blockendTheta:\depth);
\draw[densely dotted, blue] (0,0) -- (\blockstartX, \blockstartY);
\draw[densely dotted, blue] (0,0) -- (\blockendX, \blockendY);
}

\begin{tikzpicture}
  [>=latex, scale=2.8]

  \tikzset{
    pics/carc/.style args={#1:#2:#3}{
      code={
        \draw[pic actions] (#1:#3) arc(#1:#2:#3);
      }
    }
  }

  \newcommand{\axislength}{1.4};
  \newcommand{\cylinderradius}{1.2};
  \newcommand{\pointoriginal}{1.3};
  \newcommand{\pointmoved}{0.45};
  \newcommand{\blockrange}{6};

  \coordinate (O) at (0, 0);
  \coordinate (axishor) at (\axislength, 0);
  \coordinate (axisver) at (0, \axislength);
  \coordinate (p) at (\cylinderradius,\pointoriginal);
  \coordinate (pm) at (\cylinderradius,\pointmoved);
  \coordinate (rintersec) at (\cylinderradius, 0);

  \draw[->] (-0.1,0) -- (axishor) node[right] {$r$};
  \draw[->] (0,-0.1) -- (axisver) node[above] {$z$};

  \node[anchor=north east] at (O) {$o$};

  \draw (1, 0) arc (0:90:1);

  \draw[densely dotted] (\cylinderradius,-0.1) -- (\cylinderradius,\axislength);

  \fill (p) circle [radius=0.015] node[right] {$p$};
  \draw (O) -- (p) node[midway, anchor = south east, inner sep = 1pt, shift={(-2pt, 2pt)]}] {$d$};
  \setAngleTheta{\cylinderradius}{\pointoriginal}
  \drawBlockRadial{\cylinderradius}{\angleTheta}{\blockrange}
  \drawBlockCylinder{\cylinderradius}{\angleTheta}{\blockrange}
  \coordinate (label_s) at (\angleTheta:1);

  \fill (pm) circle [radius=0.015] node[right] {$p_\text{m}$};
  \draw (O) -- (pm);
  \pgfmathsetmacro\lengthl{\pointoriginal - \pointmoved}
  \drawBlockRadialMoved{\cylinderradius}{\angleTheta}{\blockrange}{\lengthl}
  \drawBlockCylinderMoved{\cylinderradius}{\angleTheta}{\blockrange}{\lengthl}
  \setAngleTheta{\cylinderradius}{\pointmoved}
  \fill (\angleTheta:1) circle [radius=0.015] node[right] {$s_\text{m}$};

  \draw pic["$\theta$", draw, angle radius=1.44cm, angle eccentricity=1.14] {angle = p--O--axisver};
  \draw pic["$\theta_\text{m}$", draw, angle radius=1cm, angle eccentricity=1.2, pic text options={shift={(-3pt,2pt)}}] {angle = pm--O--axisver};
  \draw pic["$\Delta\theta$", draw, angle radius=1.66cm, angle eccentricity=1.2] {angle = pm--O--p};
  \draw pic[pic text=., draw, angle radius=0.5cm, angle eccentricity=0.5] {angle = pm--rintersec--O};

  \fill (label_s) circle [radius=0.015] node[right, shift={(-0.75pt, 2pt)}] {$s$};

  \coordinate (p_label) at ($(p) + (0.2, 0)$);
  \coordinate (pm_label) at ($(pm) + (0.2, 0)$);
  \draw[|-|] (p_label) -- (pm_label) node[midway, anchor=east] {$l$};
  \path (O) -- (rintersec) node[midway, anchor=north] {$r$};

  \draw[thick, black, ->] ($(p) + (0, -0.25)$) -- ($(pm) + (0, 0.25)$);

\end{tikzpicture}}%
  \hfill%
  \subfloat[Geodesic geometry, backward pass\label{fig:geometry:bwd}]{
\newcommand\setAngleTheta[2]{
\pgfmathsetmacro\angleTheta{atan(#2/#1)}
}

\newcommand\drawBlockCylinder[3]{
\pgfmathsetmacro\blockstart{#2 - #3/2}
\pgfmathsetmacro\blockend{#2 + #3/2}
\pgfmathsetmacro\blockstartY{#1 * tan(\blockstart)}
\pgfmathsetmacro\blockendY{#1 * tan(\blockend)}

\draw[thick, red] (\blockstart:0.995) arc (\blockstart:\blockend:0.995);
\draw[thick, red] (#1, \blockstartY) -- (#1, \blockendY);
\draw[densely dotted, red] (0,0) -- (#1, \blockstartY);
\draw[densely dotted, red] (0,0) -- (#1, \blockendY);
}

\newcommand\drawBlockCylinderMoved[4]{
\pgfmathsetmacro\blockstart{#2 - #3/2}
\pgfmathsetmacro\blockend{#2 + #3/2}
\pgfmathsetmacro\blockstartY{#1 * tan(\blockstart) - #4}
\pgfmathsetmacro\blockendY{#1 * tan(\blockend) - #4}
\pgfmathsetmacro\blockstartTheta{atan(\blockstartY/#1)}
\pgfmathsetmacro\blockendTheta{atan(\blockendY/#1)}
\draw[thick, red] (#1, \blockstartY) -- (#1, \blockendY);
\draw[densely dotted, red] (0,0) -- (#1, \blockstartY);
\draw[densely dotted, red] (0,0) -- (#1, \blockendY);
\draw[thick, red] (\blockstartTheta:1) arc (\blockstartTheta:\blockendTheta:1);
}

\newcommand\drawBlockRadialMovedBack[4]{
\pgfmathsetmacro\blockstart{#2 - #3/2}
\pgfmathsetmacro\blockend{#2 + #3/2}
\pgfmathsetmacro\depth{#1 / cos(#2)}
\pgfmathsetmacro\yshift{-#4}
\pgfmathsetmacro\blockstartX{\depth * cos(\blockstart)}
\pgfmathsetmacro\blockstartY{\depth * sin(\blockstart) - #4}
\pgfmathsetmacro\blockendX{\depth * cos(\blockend)}
\pgfmathsetmacro\blockendY{\depth * sin(\blockend) - #4}
\pgfmathsetmacro\blockstartTheta{atan(\blockstartY/\blockstartX)}
\pgfmathsetmacro\blockendTheta{atan(\blockendY/\blockendX)}

\pgfmathsetmacro\blockcenterY{#1 * tan(#2) - #4}
\pgfmathsetmacro\blockmovedTheta{atan(\blockcenterY/#1)}

\pgfmathsetmacro\depth{#1 / cos(\blockmovedTheta)}
\pgfmathsetmacro\blockstartX{\depth * cos(\blockstartTheta)}
\pgfmathsetmacro\blockstartY{\depth * sin(\blockstartTheta) + #4}
\pgfmathsetmacro\blockendX{\depth * cos(\blockendTheta)}
\pgfmathsetmacro\blockendY{\depth * sin(\blockendTheta) + #4}
\draw[very thick, red] (\blockstartTheta:0.99) arc (\blockstartTheta:\blockendTheta:0.99);
\draw[very thick, red] (\blockstartTheta:\depth) arc (\blockstartTheta:\blockendTheta:\depth);
\draw[densely dashed, red] (0,0) -- (\blockstartTheta:\depth);
\draw[densely dashed, red] (0,0) -- (\blockendTheta:\depth);
\draw[very thick, red, shift={(0, #4)}] (\blockstartTheta:\depth) arc (\blockstartTheta:\blockendTheta:\depth);
\draw[densely dashed, red] (0,0) -- (\blockstartX, \blockstartY);
\draw[densely dashed, red] (0,0) -- (\blockendX, \blockendY);
\pgfmathsetmacro\blockstartTheta{atan(\blockstartY/\blockstartX)}
\pgfmathsetmacro\blockendTheta{atan(\blockendY/\blockendX)}
\draw[very thick, red] (\blockstartTheta:0.9875) arc (\blockstartTheta:\blockendTheta:0.9875);
}

\newcommand\drawBlockCylinderMovedBack[5]{
\pgfmathsetmacro\blockstart{#2 - #3/2}
\pgfmathsetmacro\blockend{#2 + #3/2}
\pgfmathsetmacro\blockstartY{#1 * tan(\blockstart) - #4}
\pgfmathsetmacro\blockendY{#1 * tan(\blockend) - #4}
\pgfmathsetmacro\blockstartTheta{atan(\blockstartY/#1)}
\pgfmathsetmacro\blockendTheta{atan(\blockendY/#1)}

\draw[thick, black] (\blockstart:1) arc (\blockstart:\blockend:1);

\draw[very thick, blue] (\blockstartTheta:1.01) arc (\blockstartTheta:\blockendTheta:1.01);
\draw[densely dashed, blue] (0,0) -- (#1, \blockstartY);
\draw[densely dashed, blue] (0,0) -- (#1, \blockendY);
\draw[very thick, blue] (#1, \blockstartY) -- (#1, \blockendY);

\pgfmathsetmacro\blockstartBackY{\blockstartY + #4}
\pgfmathsetmacro\blockendBackY{\blockendY + #4}
\pgfmathsetmacro\blockstartBackTheta{atan(\blockstartBackY/#1)}
\pgfmathsetmacro\blockendBackTheta{atan(\blockendBackY/#1)}
\draw[very thick, blue] (#1, \blockstartBackY) -- (#1, \blockendBackY);
\draw[densely dashed, black] (0,0) -- (#1, \blockstartBackY);
\draw[densely dashed, black] (0,0) -- (#1, \blockendBackY);
\draw[very thick, blue] (\blockstartBackTheta:1.0125) arc (\blockstartBackTheta:\blockendBackTheta:1.0125);
}

\begin{tikzpicture}
  [>=latex, scale=2.8]

  \tikzset{
    pics/carc/.style args={#1:#2:#3}{
      code={
        \draw[pic actions] (#1:#3) arc(#1:#2:#3);
      }
    }
  }

  \newcommand{\axislength}{1.4};
  \newcommand{\cylinderradius}{1.2};
  \newcommand{\pointoriginal}{1.3};
  \newcommand{\pointmoved}{0.45};
  \newcommand{\blockrange}{6};

  \coordinate (O) at (0, 0);
  \coordinate (axishor) at (\axislength, 0);
  \coordinate (axisver) at (0, \axislength);
  \coordinate (p) at (\cylinderradius,\pointoriginal);
  \coordinate (pm) at (\cylinderradius,\pointmoved);
  \coordinate (rintersec) at (\cylinderradius, 0);

  \draw[->] (-0.1,0) -- (axishor) node[right] {$r$};
  \draw[->] (0,-0.1) -- (axisver) node[above] {$z$};

  \node[anchor=north east] at (O) {$o$};

  \draw (1, 0) arc (0:90:1);

  \draw[densely dotted] (\cylinderradius,-0.1) -- (\cylinderradius,\axislength);

  \fill (p) circle [radius=0.015] node[right] {$p$};
  \draw (O) -- (p) node[midway, anchor = south east, inner sep = 1pt, shift={(-2pt, 2pt)]}] {$d$};
  \setAngleTheta{\cylinderradius}{\pointoriginal}
  \coordinate (label_s) at (\angleTheta:1);

  \fill (pm) circle [radius=0.015] node[right] {$p_\text{m}$};
  \draw (O) -- (pm);
  \pgfmathsetmacro\lengthl{\pointoriginal - \pointmoved}
  \drawBlockRadialMovedBack{\cylinderradius}{\angleTheta}{\blockrange}{\lengthl}
  \drawBlockCylinderMovedBack{\cylinderradius}{\angleTheta}{\blockrange}{\lengthl}
  \setAngleTheta{\cylinderradius}{\pointmoved}
  \setAngleTheta{\cylinderradius}{\pointmoved}
  \fill (\angleTheta:1) circle [radius=0.015] node[right] {$s_\text{m}$};

  \draw pic["$\theta$", draw, angle radius=1.44cm, angle eccentricity=1.14] {angle = p--O--axisver};
  \draw pic["$\theta_\text{m}$", draw, angle radius=1cm, angle eccentricity=1.2, pic text options={shift={(-3pt,2pt)}}] {angle = pm--O--axisver};
  \draw pic["$\Delta\theta$", draw, angle radius=1.66cm, angle eccentricity=1.2] {angle = pm--O--p};
  \draw pic[pic text=., draw, angle radius=0.5cm, angle eccentricity=0.5] {angle = pm--rintersec--O};

  \fill (label_s) circle [radius=0.015] node[right, shift={(-0.75pt, 2pt)}] {$s$};

  \coordinate (p_label) at ($(p) + (0.2, 0)$);
  \coordinate (pm_label) at ($(pm) + (0.2, 0)$);
  \draw[|-|] (p_label) -- (pm_label) node[midway, anchor=east] {$l$};
  \path (O) -- (rintersec) node[midway, anchor=north] {$r$};

  \draw[thick, black, <-] ($(p) + (0, -0.25)$) -- ($(pm) + (0, 0.25)$);

\end{tikzpicture}}%
  \caption{(a) Epipole-oriented coordinate system resulting from translational camera motion. (b) Geodesic geometry, where red elements are exclusive to the original geodesic motion model and blue elements are exclusive to the geometry-corrected geodesic motion model. (c) Visualization of effects of reverse camera motion on both motion model variants.}%
  \label{fig:geometry}%
\end{figure*}

\section{Geodesic Motion Modeling}\label{sec:groundwork}

Consider an object at a position $\vec{p} = (x, y, p) \in \mathbb{R}^3$ in 3D space that is being captured by a 360-degree camera moving along a known camera motion vector $\vec{q} \in \mathbb{R}^3$ as visualized in Fig.~\ref{fig:geometry:sphere}.
Without loss of generality, we assume $\lVert \vec{q} \rVert_2 = 1$.
The origin of the coordinate system is fixed at the camera viewpoint and is oriented such that its $z$-axis points in the direction of the camera motion vector $\vec{q}$.
Translational motion of the camera by $l \cdot \vec{q}$ results in the object position $\vec{p}$ being shifted to
\begin{align}
  \vec{p}_\text{m} = (x, y, p_\text{m})^T = (x, y, p - l)^T.\label{eq:pos_motion}
\end{align}

The 360-degree camera, which does not observe depth, perceives the object positions $\vec{p}$ and $\vec{p}_\text{m}$ through their projections $\vec{s} \in \mathcal{S}$ and $\vec{s}_\text{m} \in \mathcal{S}$ on the unit sphere, where $\mathcal{S} = \{ \vec{s} \in \mathbb{R}^3\ |\ \lVert \vec{s} \rVert_2 = 1 \}$.
The projections $\vec{s}$ and $\vec{s}_\text{m}$ can also be represented in spherical coordinates as $(\theta, \varphi)$ and $(\theta_\text{m}, \varphi_\text{m})$, respectively, where the spherical radius $\rho=1$ is omitted for clarity.
As shown in~\eqref{eq:pos_motion}, the translational motion of the camera only affects the $z$-coordinate of the object position $\vec{p}$, such that only the polar angle $\theta$ of the projected position $\vec{s}$ changes.
This follows directly from the relation between spherical and cartesian coordinates.
The resulting trajectory of the projected position $\vec{s}$ along varying polar angle $\theta$ but constant azimuth $\varphi$ denotes the so-called \textit{geodesic}.

In~\cite{Vishwanath2022}, \textit{Vishwanath et al.} utilized this knowledge to design a geodesic motion model for translational camera motion.
To model the rate of displacement along the geodesics, i.e., along~$\theta$, they assume the geometry in Fig.~\ref{fig:geometry:fwd}.
It shows a partial profile along the geodesic from Fig.~\ref{fig:geometry:sphere}, where the $z$-axes of Fig.~\ref{fig:geometry:sphere} and (b) coincide and the cylindrical radius~$r$ represents the distance of a point from the $z$-axis (cf. cylindrical coordinates).
Applying the law of sines to the triangle $o$-$p$-$p_\text{m}$ yields

\begin{align}
  \frac{d}{l} &= \frac{\sin(\pi - \theta_\text{m})}{\sin(\Delta\theta)} = \frac{\sin(\theta + \Delta\theta)}{\sin(\Delta\theta)}. \label{eq:vm_base}
\end{align}
With a given motion vector $\vec{t} = (t_u, t_v) \in \mathbb{R}^2$, the unknown fraction $k = d/l$ is calculated for the block center $(\theta_\text{c}, \varphi_\text{c})$ as
\begin{align}
  k &= \frac{d}{l} = \frac{\sin(\theta_\text{c} + \Delta \cdot t_u)}{\sin(\Delta \cdot t_u)},
\end{align}
which calculates the fraction $k$ for the scenario that the block center is displaced by $\Delta \cdot t_u$ along its geodesic.
$\Delta$ describes an angular shift per unit of motion vector.

As \textit{Vishwanath et al.} assume that all pixels $(i, j)$ in a block share the same camera displacement $l$ and depth $d$, the fraction $k$ is constant for all pixels in a block.
The rate of displacement for the remaining block pixels is then calculated by solving~\eqref{eq:vm_base} for $\Delta \theta$ yielding
\begin{align}
  \Delta\theta_{ij} = \arctan\left(\frac{\sin(\theta_{ij})}{k - \cos(\theta_{ij})}\right).
\end{align}
The resulting motion model is finally described through
\begin{align}
  \theta_{\text{m},ij} = \theta_{ij} + \Delta \theta_{ij},\quad
  \varphi_{\text{m},ij} = \varphi_{ij} + \Delta \cdot t_v. \label{eq:ged_vm}
\end{align}
The displacement along $\theta$ accounts for the camera motion along the described geodesics, the displacement along $\varphi$ accounts for potentially unrelated object motion.

\section{Geometry-corrected Geodesic Motion Modeling}\label{sec:proposal}

The assumption that all pixels in a block share the same spherical depth $d$ has a drawback.
After the pixels have been moved parallel to the camera motion vector, the spherical depth $d$ of the moved block pixels is not identical anymore, as visible in red in Fig.~\ref{fig:geometry:fwd}.
In a next time step the model will, however, again consider all pixels in a block to share the same spherical depth $d$, as visible in red in Fig.~\ref{fig:geometry:bwd}.
The interpreted geometry of the regarded object changes, which introduces undesired discontinuities into the motion modeling procedure and ultimately leads to errors when reversing the camera motion from the previous time step, as Fig.~\ref{fig:geometry:bwd} shows.

In our proposed geometry-corrected geodesic motion model, we prevent these errors by projecting the pixels onto the same manifold that is also used to represent the geodesic displacement.
This corresponds to the surface of a cylinder in 3D space which is oriented such that its $z$-axis coincides with the camera motion vector $\vec{q}$.

According to the blue elements in Fig.~\ref{fig:geometry:fwd}, we model this geometry by assuming that all pixels within a block share the same cylindrical radius~$r$.
The rate of displacement along $\theta$ is modeled by projecting both the original position $p_{ij}$ and the moved position $p_{\text{m},ij}$ onto the cylinder surface according to
\begin{align}
  p_{ij} &= r \cdot \cot(\theta_{ij}),\label{eq:cyl_p}\\
  p_{\text{m},ij} &= r \cdot \cot(\theta_{\text{m},ij})\label{eq:cyl_pm}.
\end{align}
Following~\eqref{eq:pos_motion}, $p_{ij}$ and $p_{\text{m},ij}$ are related by
\begin{align}
  p_{\text{m},ij} = p_{ij} - l = p_{ij} - \Delta_z \cdot t_u,\label{eq:cyl_mv}
\end{align}
where the displacement $l$ along the height of the cylinder is modeled by the motion vector component $t_u$.
The scaling factor~$\Delta_z$ refers to the shift along the $z$-axis that yields an angular shift equivalent to~$\Delta$, and is calculated as $\Delta_z = \tan(\Delta)$.
Solving~\eqref{eq:cyl_pm} for $\theta_{\text{m},ij}$, and inserting~\eqref{eq:cyl_mv} and~\eqref{eq:cyl_p} yields
\begin{align}
  \theta_{\text{m}, ij} = \arccot\left(\cot(\theta_{ij}) - \frac{\Delta_z \cdot t_u}{r}\right).\label{eq:cyl_ged}
\end{align}
For the cylindrical radius $r$, we allow two scaling configurations according to
\begin{align}
  r = \begin{cases}
        1 & \text{scaling = global}, \\
        \sin(\theta_c) & \text{scaling = local}.
  \end{cases}
\end{align}
With global scaling, all blocks share the same scale at the cost of blocks closer to the poles experiencing significantly reduced motion magnitudes compared to blocks close to the equator.
With local scaling, the reduction of motion magnitudes is counteracted by matching the cylindrical radius for all pixels in the regarded block with the cylindrical radius of the block center, at the cost of discontinuities in motion characteristics among different blocks.
We recommend to use global scaling and validate this choice compared to local scaling in our experiments.

Finally, the proposed geometry-corrected geodesic motion model is summarized as
\begin{align}
  \theta_{\text{m},ij} =~\eqref{eq:cyl_ged},\quad
  \varphi_{\text{m},ij} = \varphi_{ij} + \Delta \cdot t_v. \label{eq:ged_gc}
\end{align}
As shown in blue in Fig.~\ref{fig:geometry:bwd}, our geometry correction successfully eliminates the discontinuities and errors observed for the original geodesic motion model.

\section{Per-Frame Camera Motion Coding}\label{sec:camera_motion_coding}

Knowledge on the camera motion vector $\vec{q}$ is necessary for all variants of the geodesic motion model.
In~\cite{Vishwanath2022}, the camera motion vector was known beforehand and globally assigned for a given sequence.
As in realistic scenarios, the camera motion is not known beforehand and might change gradually over time, we propose to estimate per-frame camera motion at the encoder and transmit it to the decoder as part of the compressed bitstream.

In our investigations, we estimate per-frame camera motion using the classical 8-point algorithm (8PA)~\cite{Longuet-higgins1981, Hartley1997a} and
finetune the result of 8PA using dense optical flow fields according to~\cite{Pathak2017}.
Optical flow is obtained through classical (TVL1~\cite{Zach2007}, ILK~\cite{LeBesnerais2005}) and learning-based (RAFT~\cite{Teed2020}, GMA~\cite{Jiang2021}) methods.

For the resulting per-frame camera motion information, we use a predictive coding scheme.
The camera motion vector components are represented in fixed precision format with 24 fractional bits.
The camera motion vector for the current frame $\vec{q}$ is predicted as the camera motion vector $\hat{\vec{q}}$ of the nearest neighbor from the set of available frames at the decoder.
If two frames in the set share the same distance to the current frame, their camera motion vectors are averaged to create the prediction.
The difference between the actual and the predicted camera motion vector $\tilde{\vec{q}} = \vec{q} - \hat{\vec{q}}$ is then encoded in spherical domain representation $(\tilde{\theta}_\text{q}, \tilde{\varphi}_\text{q})$ and transmitted in the picture header that precedes each frame in the bitstream.
Exp-Golomb coding~\cite{Teuhola1978} is used to signal $\lvert\tilde{\theta}_\text{q}\rvert$ and $\lvert\tilde{\varphi}_\text{q}\rvert$.
The parameter $k$ is optimized for the occurring data distribution to achieve minimum average codeword length as $k=18$.
The signs of $\tilde{\theta}_\text{q}$ and $\tilde{\varphi}_\text{q}$ are signaled separately through flags.

\section{Performance Evaluation}\label{sec:performance}

\begin{table*}[t]
  \centering
  \setcounter{table}{2}
  \footnotesize
  \begin{tabular}{l||r|r|r|r||r|r|r|r||r|r|r|r}
 & \multicolumn{4}{c||}{GED+vm~\cite{Vishwanath2022}} & \multicolumn{4}{c||}{GED+gcg (ours)} & \multicolumn{4}{c}{GED+gcl (ours)} \\
\multicolumn{1}{r||}{Optical Flow Method} & TVL1 & ILK & RAFT & GMA & TVL1 & ILK & RAFT & GMA & TVL1 & ILK & RAFT & GMA \\
\hline
SkateboardInLot & -0.98 & -1.31 & -1.42 & -1.48 & \bfseries -1.22 & \bfseries -1.88 & \bfseries -1.97 & \bfseries -1.98 & -1.07 & -1.74 & -1.71 & -1.75 \\
ChairliftRide & -1.96 & -2.01 & -2.19 & -2.12 & \bfseries -2.76 & \bfseries -2.80 & \bfseries -2.83 & \bfseries -2.83 & -2.24 & -2.19 & -2.34 & -2.24 \\
Balboa & -1.97 & -1.98 & -1.78 & -1.93 & -2.20 & -2.11 & -2.07 & -2.08 & \bfseries -2.26 & \bfseries -2.21 & \bfseries -2.19 & \bfseries -2.17 \\
Broadway & -1.15 & -1.43 & -1.46 & -1.36 & -1.39 & -1.74 & -1.65 & -1.62 & \bfseries -1.44 & \bfseries -1.80 & \bfseries -1.68 & \bfseries -1.69 \\
Landing2 & \bfseries -3.45 & \bfseries -3.86 & \bfseries -3.97 & \bfseries -4.21 & -3.05 & -3.70 & -3.45 & -3.92 & -3.04 & -3.64 & -3.61 & -3.92 \\
BranCastle2 & -1.15 & -1.04 & -1.13 & -1.16 & \bfseries -1.21 & \bfseries -1.06 & \bfseries -1.17 & \bfseries -1.19 & -1.15 & -0.99 & -1.12 & -1.14 \\
\hline
Average & -1.78 & -1.94 & -1.99 & -2.04 & \bfseries -1.97 & \bfseries -2.22 & \bfseries -2.19 & \bfseries -2.27 & -1.87 & -2.09 & -2.11 & -2.15 \\
\hline
Average (w/o camera motion bits) & -1.83 & -1.99 & -2.04 & -2.09 & \bfseries -2.02 & \bfseries -2.27 & \bfseries -2.24 & \bfseries -2.32 & -1.92 & -2.15 & -2.16 & -2.20 \\

\end{tabular}
  \caption{Rate savings for geodesic motion modeling variants with 8PA and optical flow finetuning in \% with respect to VTM-17.2 based on WS-PSNR. Bold entries mark the highest rate savings for each optical flow method.}
  \label{tab:ged_optflow}
\end{table*}

We evaluate the performance of the proposed geometry-corrected geodesic motion modeling and per-frame camera motion coding concepts based on their integration into the VVC reference software VTM-17.2~\cite{Browne2022, VTM-17.2}.
The decision between geodesic motion modeling and existing inter prediction tools such as merge mode, translational or affine motion modeling is part of rate distortion optimization at the encoder\footnote{The source code of our VTM integration is publicly available at \textit{https://github.com/FAU-LMS/vvc-extension-ged}}.

Coding is performed on six sequences from the JVET 360-degree test set~\cite{Hanhart2018} at a resolution of $2048 \times 1024$ pixels.
For each sequence, 32 frames are coded in random access configuration using four quantization parameters QP $\in \{22,27,32,37\}$~\cite{Hanhart2018, Bossen2020}.
Rate savings are calculated according to the Bjøntegaard Delta (BD) model~\cite{Bjontegaard2001} with respect to the baseline VTM-17.2 based on WS-PSNR~\cite{Sun2017} as calculated by the 360Lib software 360Lib-13.1~\cite{360Lib-13.1, Ye2020a}.

\begin{table}[t]
  \centering
  \setcounter{table}{0}
  \footnotesize
  \begin{tabular}{l||r|r|r}
 & \makecell{GED+orig\\\cite{Vishwanath2022}} & \makecell{GED+gcg\\(ours)} & \makecell{GED+gcl\\(ours)} \\
\hline
SkateboardInLot & -1.33 & \bfseries -1.85 & -1.65 \\
ChairliftRide & -2.08 & \bfseries -2.70 & -2.23 \\
Balboa & -1.89 & -2.17 & \bfseries -2.27 \\
Broadway & -1.47 & \bfseries -1.76 & -1.74 \\
Landing2 & \bfseries -3.86 & -3.79 & -3.76 \\
BranCastle2 & -1.08 & \bfseries -1.14 & -1.08 \\
\hline
Average & -1.95 & \bfseries -2.23 & -2.12 \\
\end{tabular}

  \caption{Rate savings in \% for geodesic motion modeling variants with global camera motion information with respect to VTM-17.2 based on WS-PSNR. Bold entries mark the highest rate savings.}
  \label{tab:ged_flavors}
\end{table}

Table~\ref{tab:ged_flavors} shows a comparison of the different geodesic motion modeling variants with global camera motion as assumed in prior works.
The original geodesic motion modeling variant by \textit{Vishwanath et al.}~\cite{Vishwanath2022} is denoted as GED+orig, the proposed geometry-corrected variants with global and local scaling are denoted as GED+gcg and GED+gcl, respectively.
The geometry-corrected variants yield significant rate savings over GED+orig in both scaling configurations.
On average, GED+gcg and GED+gcl achieve rate savings of 2.23\% and 2.12\% outperforming GED+orig by 0.28 pp~(percentage points) and 0.17 pp.

Table~\ref{tab:ged_complexity} shows a comparison of the required number of computational operations for the different geodesic motion modeling variants per block of size $M \times N$.
The presented values follow directly from~\eqref{eq:pos_motion} -~\eqref{eq:ged_vm} and~\eqref{eq:cyl_p} -~\eqref{eq:ged_gc}.
TRIG denotes trigionometric function evaluations, MUL denotes multiplications, DIV denotes divisions, and ADD denotes additions.
Both proposed geometry-corrected variants show a decreased complexity compared to GED+orig, where especially the amount of costly TRIG operations is reduced.
The recommended global scaling proves to be favorable in terms of both achieved rate savings and computational complexity.

Table~\ref{tab:ged_optflow} shows the rate savings achieved by the different geodesic motion modeling variants with the proposed per-frame camera motion.
The camera motion is estimated using 8PA, finetuned using different dense optical flow fields, and encoded in the frame headers according to Section~\ref{sec:camera_motion_coding}.
With the exception of TVL1, both geometry-corrected motion modeling variants surpass the rate savings achieved for GED+orig in the global camera motion information scenario.
With rate savings of 2.27\% on average, GED+gcg with GMA performs best in the field and outperforms GED+orig with GMA by 0.23 pp on average.
The proposed per-frame camera motion information shows clear benefits especially for GED+gcg with GMA, which outperforms all geodesic motion modeling variants with global camera motion information.
The prior state-of-the-art GED+orig without per-frame camera motion information is surpassed by 0.32 pp.
Global scaling continues to be favorable over local scaling.

The average rate savings \textit{without camera motion bits} quantify the maximum rate savings that could be achieved using the estimated camera motion information without having to signal any camera motion information to the decoder and represent an upper bound for the achievable rate savings.

\begin{table}[t]
  \centering
  \footnotesize
  \begin{adjustbox}{width=\linewidth,center}
  \begin{tabular}{l||l|l|l|l||l}
  & \makecell[c]{TRIG} & \makecell[c]{MUL} & \makecell[c]{DIV} & \makecell[c]{ADD} & \makecell[c]{TOTAL} \\
  \hline
  GED+orig & $3MN + 2$   & $1$   & $1MN + 1$   & $2MN + 1$   & $6MN + 5$     \\
  GED+gcg & $2MN$ & $1MN$ & $0$ & $1MN$ & $4MN$ \\
  GED+gcl & $2MN + 1$ & $1MN$ & $1MN$ & $1MN$ & $5MN + 1$ \\
\end{tabular}

  \end{adjustbox}
  \caption{Comparison of the complexity of the different geodesic motion modeling variants per block of size $M \times N$.}
  \label{tab:ged_complexity}
\end{table}

\section{Conclusion}\label{sec:conclusion}

In this paper, we proposed a geometry-corrected geodesic motion model that enhances the state of the art by representing object positions and camera motion displacements on the same manifold.
We represent the according manifold as a cylinder in 3D space that is oriented along a known camera motion vector.
We additionally propose coding of per-frame camera motion information using a predictive coding scheme employing exp-golomb coding.
In our experiments, camera motion is estimated at the encoder using a combination of the classical eight point algorithm and optical flow field finetuning.
Our geometry-corrected geodesic motion model with estimated per-frame camera motion achieves average rate savings of up to 2.27\% over H.266/VVC, representing an improvement of more than 0.32 pp over the original geodesic motion model despite reduced computational complexity.

\bibliographystyle{IEEEbib}
\bibliography{ms}

\begin{thebibliography}{10}

\bibitem{Bross2021a}
Benjamin Bross, Ye-Kui Wang, Yan Ye, Shan Liu, Jianle Chen, Gary~J. Sullivan,
  and Jens-Rainer Ohm,
\newblock ``Overview of the {{Versatile Video Coding}} ({{VVC}}) {{Standard}}
  and its {{Applications}},''
\newblock {\em IEEE Trans. Circuits Syst. Video Technol.}, vol. 31, no. 10, pp.
  3736--3764, Oct. 2021.

\bibitem{Pearson1990}
I.~I. Pearson,
\newblock {\em Map {{Projections}}: {{Theory}} and {{Applications}}},
\newblock {CRC Press}, {Boca Raton, Fla}, 2nd edition edition, Mar. 1990.

\bibitem{Wien2017}
Mathias Wien, Vittorio Baroncini, Philippe Hanhart, Jill Boyce, Andrew Segall,
  and Teruhiko Suzuki,
\newblock ``Results of the {{Call}} for {{Evidence}} on {{Video Compression}}
  with {{Capability}} beyond {{HEVC}}, {{JVET-G1004}},''
\newblock in {\em Proc. 7th {{Meet}}. {{Jt}}. {{Video Explor}}. {{Team}}}, July
  2017, pp. 1--17.

\bibitem{Li2019}
Li~Li, Zhu Li, Xiang Ma, Haitao Yang, and Houqiang Li,
\newblock ``Advanced {{Spherical Motion Model}} and {{Local Padding}} for
  360{\textdegree} {{Video Compression}},''
\newblock {\em IEEE Trans. Image Process.}, vol. 28, no. 5, pp. 2342--2356, May
  2019.

\bibitem{DeSimone2017}
Francesca De~Simone, Pascal Frossard, Neil Birkbeck, and Balu Adsumilli,
\newblock ``Deformable {{Block-Based Motion Estimation}} in {{Omnidirectional
  Image Sequences}},''
\newblock in {\em Proc. {{IEEE}} 19th {{Int}}. {{Workshop Multimed}}. {{Signal
  Process}}.}, Oct. 2017, pp. 1--6.

\bibitem{Vishwanath2017}
Bharath Vishwanath, Tejaswi Nanjundaswamy, and Kenneth Rose,
\newblock ``Rotational {{Motion Model}} for {{Temporal Prediction}} in 360
  {{Video Coding}},''
\newblock in {\em Proc. {{IEEE}} 19th {{Int}}. {{Workshop Multimed}}. {{Signal
  Process}}.}, Oct. 2017, pp. 1--6.

\bibitem{Vishwanath2022}
Bharath Vishwanath, Tejaswi Nanjundaswamy, and Kenneth Rose,
\newblock ``A {{Geodesic Translation Model}} for {{Spherical Video
  Compression}},''
\newblock {\em IEEE Trans. Image Process.}, vol. 31, pp. 2136--2147, Feb. 2022.

\bibitem{Longuet-higgins1981}
H.~C. {Longuet-Higgins},
\newblock ``A {{Computer Algorithm}} for {{Reconstructing}} a {{Scene}} from
  {{Two Projections}},''
\newblock {\em Nature}, vol. 293, no. 5828, pp. 133--135, Sept. 1981.

\bibitem{Hartley1997a}
Richard Hartley,
\newblock ``In {{Defense}} of the {{Eight-Point Algorithm}},''
\newblock {\em IEEE Trans. Pattern Anal. Mach. Intell.}, vol. 19, no. 6, pp.
  580--593, June 1997.

\bibitem{Pathak2017}
Sarthak Pathak, Alessandro Moro, Hiromitsu Fujii, Atsushi Yamashita, and Hajime
  Asama,
\newblock ``Spherical {{Video Stabilization}} by {{Estimating Rotation}} from
  {{Dense Optical Flow Fields}},''
\newblock {\em J. Robot. Mechatron.}, vol. 29, no. 3, pp. 566--579, June 2017.

\bibitem{Zach2007}
C.~Zach, T.~Pock, and H.~Bischof,
\newblock ``A {{Duality Based Approach}} for {{Realtime TV-L1 Optical Flow}},''
\newblock in {\em Proc. 29th {{DAGM Conf}}. {{Pattern Recognit}}.}, Sept. 2007,
  pp. 214--223.

\bibitem{LeBesnerais2005}
Guy Le~Besnerais and Fr{\'e}d{\'e}ric Champagnat,
\newblock ``Dense {{Optical Flow}} by {{Iterative Local Window
  Registration}},''
\newblock in {\em {{IEEE Int}}. {{Conf}}. {{Image Process}}.}, Sept. 2005,
  vol.~1, pp. I--137.

\bibitem{Teed2020}
Zachary Teed and Jia Deng,
\newblock ``{{RAFT}}: {{Recurrent All-Pairs Field Transforms}} for {{Optical
  Flow}},''
\newblock in {\em Proc. {{Eur}}. {{Conf}}. {{Comput}}. {{Vis}}.}, Aug. 2020,
  pp. 402--419.

\bibitem{Jiang2021}
Shihao Jiang, Dylan Campbell, Yao Lu, Hongdong Li, and Richard Hartley,
\newblock ``Learning {{To Estimate Hidden Motions With Global Motion
  Aggregation}},''
\newblock in {\em Proc. {{IEEE/CVF Int}}. {{Conf}}. {{Comput}}. {{Vis}}.}, Oct.
  2021, pp. 9772--9781.

\bibitem{Teuhola1978}
Jukka Teuhola,
\newblock ``A {{Compression Method}} for {{Clustered Bit-Vectors}},''
\newblock {\em Inf. Process. Lett.}, vol. 7, no. 6, pp. 308--311, Oct. 1978.

\bibitem{Browne2022}
Adrian Browne, Yan Ye, and Seung~Hwan Kim,
\newblock ``Algorithm {{Description}} for {{Versatile Video Coding}} and {{Test
  Model}} 17 ({{VTM}} 17), {{JVET-Z2002}},''
\newblock in {\em Proc. 26th {{Meet}}. {{Jt}}. {{Video Experts Team}}}, Apr.
  2022, pp. 1--141.

\bibitem{VTM-17.2}
{JVET},
\newblock ``{{VVC Reference Software VTM-17}}.2,''
  https://vcgit.hhi.fraunhofer.de/jvet/VVCSoftware\_VTM/-/releases/VTM-17.2,
  July 2022.

\bibitem{Hanhart2018}
Philippe Hanhart, Jill Boyce, Kiho Choi, and J.L. Lin,
\newblock ``{{JVET Common Test Conditions}} and {{Evaluation Procedures}} for
  360{\textdegree} {{Video}}, {{JVET-L1012}},''
\newblock in {\em Proc. 12th {{Meet}}. {{Jt}}. {{Video Explor}}. {{Team}}},
  Oct. 2018, pp. 1--7.

\bibitem{Bossen2020}
Frank Bossen, Jill Boyce, Karsten S{\"u}hring, Xiang Li, and Vadim Seregin,
\newblock ``{{VTM Common Test Conditions}} and {{Software Reference
  Configurations}} for {{SDR Video}}, {{JVET-T2010}},''
\newblock in {\em Proc. 20th {{Meet}}. {{Jt}}. {{Video Experts Team}}}, Oct.
  2020, pp. 1--2.

\bibitem{Bjontegaard2001}
Gisle Bj{\o}ntegaard,
\newblock ``Calculation of {{Average PSNR Differences}} between {{RD-curves}},
  {{VCEG-M33}},''
\newblock in {\em Proc. 13th {{Meet}}. {{Video Coding Experts Group}}}, Mar.
  2001, pp. 1--5.

\bibitem{Sun2017}
Yule Sun, Ang Lu, and Lu~Yu,
\newblock ``Weighted-to-{{Spherically-Uniform Quality Evaluation}} for
  {{Omnidirectional Video}},''
\newblock {\em IEEE Signal Process. Lett.}, vol. 24, no. 9, pp. 1408--1412,
  Sept. 2017.

\bibitem{360Lib-13.1}
{JVET},
\newblock ``{{360Lib Software 360Lib-13}}.1,''
  https://vcgit.hhi.fraunhofer.de/jvet/360lib/-/tags/360Lib-13.1, Oct. 2021.

\bibitem{Ye2020a}
Yan Ye and Jill Boyce,
\newblock ``Algorithm {{Descriptions}} of {{Projection Format Conversion}} and
  {{Video Quality Metrics}} in {{360Lib Version}} 12, {{JVET-T2004-v2}},''
\newblock in {\em Proc. 20th {{Meet}}. {{Jt}}. {{Video Experts Team}}}, Oct.
  2020, pp. 1--65.

\end{thebibliography}

\end{document}